# Towards a Physical Theory of Subjective Mental States

*Sean Lee*
(Submitted to Foundations of Physics, Oct. 8, 2007)

**Abstract**

Any complete theory of physical reality must allow for the ubiquitous phenomenon of subjective experience at some level, or risk being conceptually incoherent. However, as long as the ontological status of subjectivity itself remains unresolved, the topic will be seen as more within the purview of philosophy than of physics. Towards a resolution of this issue within empirically motivated physical theory, this article introduces an operational definition that utilizes the general consensus that subjective mental states, whatever else is controversial about them, at least correlate in some way to physical states. It is shown here that implementing this underappreciated assumption within the framework of a physical theory in fact leads to wide-ranging consequences. In particular, a correlation requires there exist a well-defined mapping from a space of subjective mental states onto a space of information-bearing states of some physical theory. Given the peculiar nature of subjective states as inherently private appearances, any empirical identification of states must be performed by the experiencing subject. It is argued that such an operationally defined 'self-measuring' act leads unavoidably to an 'uncertainty principle' that is analogous in some intriguing ways to Heisenberg's principle for quantum mechanics. A model is then introduced for subjective states as algorithmically incomputable numbers. Additionally, an inequality similar to Bell's theorem may be derived, indicating an analogy with the violations of local reality and the ontology of observables within quantum mechanics.

## 1. INTRODUCTION

The ubiquitous phenomenon of private, subjective, mental experience has been well-known for centuries to be maddeningly difficult to approach within the context of the natural sciences, indeed within any objective system of inquiry. Controversies persist no less today than in ancient times on the most basic of questions, beginning with the very explanandum itself. An equal lack of consensus exists regarding the causal role and ontological status of subjective, conscious experience in what appears to be an otherwise physical, material world. In a rather dry assessment from *The International Dictionary of Psychology* (Sutherland, 1989) we read about consciousness:

> *It is impossible to specify what it is, what it does or why it evolved. Nothing worth reading has been written about it.*

Of course, the utter lack of any real progress since the modern formulation of the problem by Descartes (Descartes, 1988) might be a sign that the problem simply has no real solution, or at least any type of solution that our minds would recognize as such (McGinn, 1989). In fact, if the mind-body problem does turn out one day to have any resolution, it seems likely to be bound by what we can think of as the Conservation of Absurdity rule: the total amount of absurdity in any logically consistent treatment of the mind-body problem is invariant with respect to transformations of methodology!

Given the manifestly subjective and apparently ineffable nature of the subject matter, consciousness research, except in occasional connection with interpretations of quantum mechanics, has historically remained much less a topic in its own right within physics. Still, it may be said that understanding the nature and place of subjectivity within the physical world is not only the longest-standing open problem of the scientific revolution; in a way it is *the* open problem of open problems. The ultimate goal of physics, phrased in the broadest possible way, is to explain why we observe what we observe. Yet what meaning is there to an observed phenomenon without the subjective experience of observation? Objective explanations of the blueness of the sky in terms of Rayleigh scattering and the neurobiology of visual processing are satisfying up to a point, but ultimately we want an understanding of *why we see the sky as blue*. More importantly, any fundamental theory that excludes subjectivity is not only incomplete; it is ultimately unintelligible, for the simple reason that we can't meaningfully claim to conceive of a world that cannot be conceived of in the first place. So in the end, it seems necessary to understand subjectivity for a complete explanation of any natural phenomenon.

## 2. THE HARD PROBLEM

The most controversial and intractable aspect of subjectivity, and the one of ultimate interest here, is often referred to as the 'Hard Problem' (Chalmers, 1995) of consciousness – as distinct from the other so-called 'Easy Problems' of consciousness. In essence, the Easy Problems are all those which can be ultimately defined exclusively in terms of objectively observable data. This data can include everything from a subject's written and verbal utterances, images of facial and body movements, as well as fMRI, EEG scans, single neuron activity, and so on. Easy Problems also include questions of classical psychology, such as, e.g., "Why did he choose chocolate?", or "How will she react to the bad news?", because one can, at least in principle, conceive of physical, explanatory models to account for the objectively observed act of 'choosing' and 'reacting'. Choosing to define private, subjective mental states wholly in terms of what is objectively observable about them by third parties is the inspiration for the famous Turing Test (Turing, 1950) and subsequent work in artificial intelligence (AI). In particular, Turing imagined future computational models of the mind which could accurately mimic the observable output of the human mind. Such models would be considered solutions (albeit non-unique) to the Easy Problems.

By contrast, the Hard Problem, true to its name, already begins with the very attempt to state it. Because its formulation in objective terms as an unambiguous explanandum is highly nontrivial and controversial[1], it is better got at by example: imagine you stub your toe and yell "Ouch!" Explaining the complex causal chain of your reaction in physical terms is wholly an Easy Problem as described above, except for the nagging residual question: why did it actually *hurt*, indeed, why did stubbing your toe feel like anything at all? Why is information processing by nerve tissue accompanied by any private, first person sensation? Nowhere within the explanatory paradigm of physical causation, however computationally complex and rich in emergent phenomenon it may be, does the *phenomenon of your private perspective* appear. This point is often illustrated with the famous 'zombie' thought experiment (Moody, 1994): Imagine an exact physical duplicate of yourself, atom for atom, with the single exception that your duplicate doesn't *feel* anything; a 'zombie' of yourself. Being your exact physical duplicate, your zombie nevertheless behaves in ways indistinguishable from you under all conceivable physical circumstances. For example, your zombie yells "Ouch!" with the identical neuro-physiological activity after stubbing a toe as you do (and some zombies even write papers on the zombie thought experiment). In the context of Easy Problems there is no objective, measurable distinction between you and your zombie, but at the same time, by hypothesis, your zombie *feels nothing*, and has no subjective experience, whereas you do. These examples are intended to illustrate that from the point of view of any conceivable objective physical theory, subjective experience in the visceral sense of '*what it feels like*' seems unexpected, mysterious and indeed superfluous.

The Hard Problem itself is controversial, and some, such as Dennett (Dennett, 1984) and Hofstadter (Hofstadter, 1979) maintain that subjectivity is best seen not as a 'real' phenomenon, but rather as a sort of emergent illusion of complex and self-referential information processing. However, it has been countered extensively by Chalmers (Chalmers, 1996) and many others that such a proposal merely skirts the true issue. Indeed, it seems manifestly incoherent to posit that experience itself is any kind of non-experiential artifact. In a very direct sense, that fact of subjectivity seems to be the one observation we actually know with certainty. At issue in this dispute is more than semantics, but basic ontological perspective that impacts any theory of physical reality: in what sense is it meaningful to ascribe the notion of 'existence' to our private sensations and first person perspectives, and how can they be related to our (apparently quite different) notion of existence in the physical universe?

## 3. AN OPERATIONAL APPROACH

Framed in this basic way, one place to start might be to ask whether any final understanding of the mind has to be part and parcel of our overall concept of basic reality. This seemingly overreaching question is motivated by the inescapable fact that whatever conception we as human beings may have of basic reality is itself obviously and necessarily a product of the very minds we wish to explain. So in the end, it may be inevitable that the two are related at the deepest level. It should be emphasized that the motivation here is much less a metaphysical one than a pragmatic one: In particular, this paper will attempt to address the question: *what types of consistent, physical theories of subjective experience are human minds capable of generating?*

---

[1] Indeed, in some sense it may be said that finding an unambiguous, objective statement of the Hard Problem would be functionally equivalent to solving it!

This article takes a purposely 'pedestrian-physics' approach to the mind, and tries to treat it as best as possible with the same methodology as it might be treated as a natural phenomenon within fundamental physics. Specifically, this will involve a three-step process of

1. Addressing the observed phenomenon, rather than a presumed representation. The phenomenon of subjectivity at the core of the Hard Problem is precisely its private, ineffable *appearance to the subject* in a way that seems to defy objective description. Thus it is crucial to find a way to allow for subjective experience, not only as objective theoretical constructs, (e.g., as peculiar brain and/or computational states) but rather directly as it appears to the *subject*.

2. Seeking an operational definition of the phenomenon. Historically, operational definitions have been the gold standard for constructing meaningful physical theories, since they force a precise definition how an object of interest is to be implemented in the theory. Any discourse which rises to the level of a science that can be shared among scientists, independent of meta-physical assumptions, ultimately requires operational definitions for any workable framework.

3. Looking for analogies in existing theories. It seems that the mysterious phenomenon of subjectivity is such an unfamiliar entity within physical theory that any progress will likely require orientation along analogs to existing frameworks. Even if these analogies are purely formal, (as in, for example, the analogy between thermodynamics and electrical circuits) they may serve as useful guides and further inspiration for a more substantive theory.

## 4. SUBJECTIVITY AS A NATURAL PHENOMENON

If we think of the existence of the experiencing mind as an observed natural phenomenon, then the only aspect of that phenomenon that is *directly observable* is our own subjective experience. Everything else we believe about the mind is indirectly observed, or rather, inferred from behavior and other external, physical data. Indeed the zombie thought experiment derives precisely from this peculiar fact. Whether or not zombies are actually possible in this or any other world, we can only talk about them coherently because the subjectivity of one is not accessible to others for direct inspection. Of course, this is exactly what's frustrating and makes the mind-body problem so hard to deal with in the way we're used to dealing with other phenomena. Before giving in to despair, however, it's important to remember that this does not render subjectivity unobserved or unnatural, but it does mean it *a priori* lacks any 'objective' definition in this sense. Nevertheless, the goal is to take subjectivity itself seriously as a natural phenomenon, while resisting the temptation to *a priori* substitute for it a physical representation such as, e.g. brain states. How can this be done within the framework of physical theory, indeed within any formal language?

Of course, there is enormous controversy on almost every aspect of the mind-body problem, in particular on the relationship between subjective experiences and the objective physical world. But whatever is seen as their causal or explanatory connection to the physical world, and irrespective of their semantic and ontological status, there is broad consensus that subjective experiences are, if nothing else, at least unambiguously *correlated* to objective physical states. That is, subjective experience is believed to necessarily occur when certain physical conditions are also present. With the beginning of the modern scientific tradition, the activity of (specifically human) brains was commonly seen as producing the necessary and sufficient conditions for subjective experience. Indeed, the goal of consciousness research within the program of standard neurobiology is to examine these Neural Correlates of Consciousness or NCCs (Koch, 2004), although few still see subjective experience as strictly limited to humans. Alternatively, the more recent research programs of artificial intelligence (AI) and representationalism see the correlation of subjectivity to biological neural states as only one possible expression of a deeper-level correlation to the computations of a Turing machine. While it is disputed by Penrose and many others that computational complexity by itself is sufficient, it is broadly held that the salient feature for consciousness is its correlation to some form of physical information processing. The dualist school, of course, sets out from the premise that subjective and objective states are *a priori* correlated.

Thus, whatever else is controversial about subjectivity, there is general agreement that subjective state at least correlates to some sort of physical and/or informational states. Within the context of a physical science, this assumption translates into requiring an unambiguous mapping **M** between whatever is identified as the space of physical states **P** with whatever is identified as the space of subjective states **Q**. We'll call this rather underappreciated assumption the *mapping principle*:

$$M: Q \to P$$

For the purposes of such a mapping, it only matters that the spaces **P** and **Q** are well-defined, and not whether the states in **P** are intended to be ontologically 'purely physical' (as, say, in physicalism) or 'purely computational' (as, say, in AI) or some sort of combination of both. For convenience they'll all be collectively referred to from here on as 'physical'. It's important to emphasize that the requirement of a mapping is independent of whatever ontological, semantic and causal relation subjective states may or may not have with any physical states. That is, the mere claim that subjectivity is at least *correlated* with physical states implies the existence of a well-defined mapping between the two.

## 5. WHAT IS A PHYSICAL STATE?

In the context of classical physics, as well as in most current philosophical discussions on the mind-body problem[2], physical reality is presumed to have an ontological status independent of observers. With the single exception of quantum theory, this assumption has obviously had great power and practical value in constructing testable physical theories. Nevertheless, as noted above, its *a priori* acceptance might be an unnecessary limiting constraint when considering possible theories of subjectivity mental states, for the simple reason that such a theory, being itself a product of a mind which it seeks to account for, can be expected to have quite a contorted semantic and ontological foundation! Be that as it may, of interest here is what physical theories of the mind can be consistently constructed by human beings. Hence we will dispense with all such ontological assumptions. Here, the relevant physical space **P** at any given time is necessarily a human construction, the physical *theory du jour*.

Call such a theory an **E theory** if it captures those observations and data believed by the theorist or community of theorists to correlate with subjective experience. Given such an ambiguous specification, there are in principle many choices available, largely depending on which observations are considered relevant. For example, in a simple computational neural network model of the mind, each state in **P** is given by the momentary connection strengths and on/off status of all nodes in the network. Current neuroscience considers the electro-chemical activity of physical neurons to form a sufficient explanatory basis, in which case the **E theory** would entail the previously mentioned NCCs. We can radically broaden the scope of this **E theory** and still stay within its conceptual foundation by imagining all of semi-classical physics applied to the whole subject and its immediate environment. States in **P** would then be given by the distributions, momenta and valences of all semi-classical atoms within and near the organism (by the generally presumed chain of supervenience, this entails all the understood rules of thermodynamics, chemistry, biochemistry, cell biology and the like). Of course, classical and semi-classical physics (in non-relativistic settings) have long been supplanted by quantum mechanics, so in principle a purely quantum mechanical description of the physical state of the subject should be at least equally valid (if not manifestly illuminating). In this case **P** might be considered to be a space of state vectors $|\Psi\rangle$, or perhaps alternatively, a relevant subset of measured observables $\{O_1, O_2 \dots O_N\}$.

Imagining still further ahead to future physical theories, we could imagine a grand physics in the 23$^{rd}$ century with, perhaps, a consistent theory of strings, loop quantum gravity, dark matter, warp drives, time machines, etc. But whatever the choice, the point is that **P** belongs to some *theory*, and as such consists of elements within some formal language. In other words, each element $p \in \mathbf{P}$ is represented by a string of symbols, which in turn can be mapped to a real number $r \in \mathbf{R}$. So any physical state we might conceive of (literally) can be represented by a unique real number. For convenience, representations in this context will be called *codes*, **C**, and the real numbers assumed in binary.

---

[2] A significant exception is, of course, the idealist school of philosophy.

Physical states are, of course, generally represented quantitatively, but since most physical models are constructed in several dimensions (spatio-temporal and others) it is obviously unusual to express states as one dimensional binary sequences until they are coded into machine language for computer simulations and modeling. Here the conceptual purpose of coding with numerical/binary sequences is to more easily be able to apply the notions of algorithmic information theory to both physical and subjective states. Specifically, and towards the goal stated in the introduction, we'll be interested in the *algorithmic computability* of the representations of such states. In doing so however it is important to note that, at this stage, such a mapping to single binary sequences is purely a *descriptive* mode and does not further restrict the structure of the theory nor define how states evolve might in time. In particular, and in spite of using the same language as for digital computers, the use of a binary coding does not require that the time-evolution of states in **P** proceed via any set of well-defined algorithms. Thus, no condition of determinism or even causality is implied by a binary coding procedure.

## 6. OPERATIONAL DEFINITION OF SUBJECTIVE STATES

As noted, the question of just what a subjective state **q** 'is' in the sense of its first person *appearance* to us is conceptually problematic. Nevertheless, following through with the program of correlating with physical state requires similarly coding each subjective state **q** ∈ **Q** with a unique real number as well. As previously stated, it's vital to note that since we're interested in mapping and eventually understanding the relationship between subjectivity to objective physical data, *we must seek to code for the subjective state per se, and not for a presumed physical representation for that state*. Given our natural human tendency to represent by proxy that which can't otherwise be readily conceived, the temptation to code for a presumed representation (such as a particular neurological or computational state) is great. However, such a substitution in this case would clearly be circular and defeat the purpose of the exercise. Any description or coding of a subject's neural, computational or other physical state properly belongs to the physical state space **P**, and not the subjective state space **Q**, regardless of any relationship between the two that may later be posited or established.

As an *observed* phenomenon, that is, as it is actually experienced, subjectivity is inherently private. The inherently private nature of subjectivity forces us to accept that *any operationally meaningful coding must be done by the experiencing subject.* This is such an important point it bears emphasizing. One of the properties that set subjectivity apart from its external expressions is precisely that no one else other than the experiencing subject has direct access to that particular realization of the phenomenon. Unfortunately, such a coding then clearly requires, among other things, fairly high-level language and communication skills on the part of the subject(s) in question. It also requires a willingness to do such coding. Thus we come to the uncomfortable first conclusion that any meaningful operational definition that might rise to the standard of a science of subjectivity can only be given for a very limited number of subjects; namely for those that are willing and able to communicate with scientists! Initially, this seems in many ways completely unsatisfactory, because we naturally ascribe subjective experience to a much broader set of living beings; young children, language-impaired persons, other primates, cetaceans, dogs, among perhaps many, many others species.

Of course, the conclusion here is not that only eloquent and cooperative humans have subjective experiences – which I believe truly would be absurd – but the discussion highlights an important limitation in our ability to treat subjectivity at all within what we understand as a physical *science*. Namely, a human science of subjectivity requires that at least some subjects be able to communicate their experiences in a way that the scientists can understand, theorize, and ultimately have their own subjective experiences about. This by itself is not so strange; after all, given that science is a human activity, any science of any phenomenon requires that scientists be able to access and understand data considered relevant for that phenomenon. In the case of subjectivity, however, the only relevant, operationally definable data that is accessible are from subjects possessing high-level language skills that are communicable to scientists. Of course, depending on the particular theory of mind that can be constructed on this initial basis, it seems reasonable to later extrapolate what is learned from such subjects to at least some other subjects not having these abilities. In this way we would ascribe subjectivity to a broader class of subjects, depending on whatever theory eventually emerges.

However, in terms of methodology, no operationally definable procedure can justify this extrapolation: it is based solely upon our own belief that we can empathize, or share common experience with, these other subjects. We can speculate that perhaps this deep-seated belief comes from the brain's mirror neurons which let us simulate in our own minds what we believe to be the experience of others (Gallese, 2001) In the case of small children we're also helped in this by our own childhood memories. Additionally, most can probably also imagine, if only for a few moments, what it would feel like to have no language available to us, and yet still have subjective experience. Of course, our ability to simulate in our own minds that which we take to occur in other minds gets much trickier with other species. The latter point is the reason Nagel's famous question relating to the issue of subjectivity "What is it like to be a bat?" (Nagel, 1974) is such a compelling illustration of the Hard Problem. In conclusion then, we will limit our methodology of discussing operationally meaningful coding schemes for subjective experience to those subjects with high-level language and communication skills, and the willingness to use those skills.

## 7. LIMITATIONS OF CODING WITH NATURAL LANGUAGE

Of course, on a practical day-to-day level, we code our subjective experiences all the time with real numbers; it's just that those real numbers happen to be coded themselves as word-strings in some natural language. Word strings like 'I'm hungry', 'I miss that look in her eyes whenever she ate jalapeños', Hamlet's soliloquy, and so on, can all be translated into binary code (and they are as soon as I type them into my laptop.) But whether succinct or verbose, such codes can only partially convey certain aspects of our experiences. Saying 'I'm hungry' doesn't say anything about what else is going on in my mind. Thus, even with articulate and willing subjects to provide input, we still have a problem: any natural language description gives us highly *incomplete codes* for subjective experiences. To define a state space **Q**, however, we're interested in a *complete code of the entire subjective state q*. That is, a real number **C(q)** that uniquely identifies each individual, entire experience **q**.

One way to improve the accuracy of these subjective state codes would be to simply use longer descriptions to capture more information: a seven page essay on hunger would convey more data than the mere words 'I'm hungry'. Alternatively, one could imagine introducing more words into natural languages, say, with each word in the dictionary having an arbitrary numbers of 'flavors' for a more nuanced rendering of the experience of hunger: hunger(1), hunger(2),…, hunger(N), and any other word, as desired. However, while this may improve things somewhat, neither scheme is workable: few subjects will remember more than the order of a few thousand words anyway, presumably much smaller than the total number of subjective experiences to be described (however they are defined.) More importantly, introducing more words and longer combinations of words create the additional complication that they take time to access from memory and formulate, during which the original experience has long passed. Generally, we can say that such a prodigious mental feat itself would, to whatever limited extent it could be done at all, take up so much of the subject's mental resources as to render the accuracy of the output code **C(q)** highly questionable in the end. That is to say, the very mental act of trying to code an experience accurately in this way would introduce a significant 'measurement error' in the code. While this particular example appears at first glance to highlight a mere practical difficulty, in fact it is illustrative of a more principle limitation in constructing operational definitions of subjective states. Indeed it will be seen that there is a fundamental and irreducible 'measurement error' in defining subjective states.

## 8. GEDANKENEXPERIMENT: CODING IN THE 23$^{rd}$ CENTURY

The principle weaknesses of having a subject generate word-strings to describe his momentary subjective state are two-fold. First, as discussed above, the mental feat of juggling many words and memories introduces a large 'measurement error'. Equally important is the fact that any self-generated word-string will only address certain *aspects* of an experience, but not the *entire* experience itself. For example, if I look up at a clear sky on a warm day and have a certain subjective experience, one aspect of that experience I might report on is the particular *blueness* I perceive; another might be the *warmth of the breeze* I perceive on my face; yet another aspect might be the particular *itch on my nose* I perceive, or the faint memory of *what I had for lunch*, and so on. But this still does not amount to whatever entire experience I had at a certain instant. An exhaustive list of the different aspects I can report on from an experience gives a lot of information, but that still does not completely define the *entire subjective experience itself*. An entire subjective experience, then, is in a deep sense manifestly more than the mere sum of its aspects, i.e. more than reports about that experience.

In order to overcome these limitations, we'll perform the following thought experiment, designed to allow the test subject to report on *entire* experiences in *binary* and only recall the most recent memory state. Imagine a hypothetical 23$^{rd}$ century neuroscientist who has a candidate **E theory** that we'll call **E$_{23}$ theory** that needs to be tested on several subjects. Remember that **E$_{23}$ theory** defines physical states **p** ∈ **P** which, by claim, are the set of states observed by the neuroscientist to be correlated to the subjects' subjective experiences. Imagine further the neuroscientist has built a 'memory-playback' system, a device which can recreate any given physical state **p** ∈ **P** as often as needed. The task is now to use the memory-playback device and the **E$_{23}$ theory** to help the subject create a code for the experienced subjective states. In essence, the neuroscientist and the memory-playback system will serve as external memory devices for the subject.

Note that the actually numbering of these states is arbitrary in the same sense that picking a reference coordinate system for measuring distance is arbitrary; what counts for the following is the measuring procedure. Thus the coding process begins with the experimenter setting an arbitrary reference state **p$_0$**, and arbitrarily denoting the subjects' momentary subjective state as **q$_0$**. This gives a first reference pair (**p$_0$**, **q$_0$**). This pairing may be confirmed by repeatedly generating the state **p$_0$**, each time immediately afterwards asking the subject:

*'Are you **certain** that your experience now is **any different** than just a moment ago?'*

Note the question references the entire experience and asks for *certainty*, so only a Yes/No answer may be given. The subject is thus reporting, to the best of his ability on his entire experience in binary, which will be later translated into the coding of the physical states of the **E$_{23}$ theory**. Since in this instance the state **p$_0$** is repeatedly generated from a good candidate **E$_{23}$ theory**, the subject should answer '*No*', in which case the new momentary subjective state, now arbitrarily labeled **q′,** will be said to be in the same *q-equivalence class* **q̣** as **q$_0$**, written **q′**≈ **q$_0$** ∈ **q̣**. By definition, if the **E$_{23}$ theory** is adequate, these results will be repeatable. (The need for such an equivalence class, as opposed to merely setting **q′**=**q$_0$** identically, will be described in the next section.)

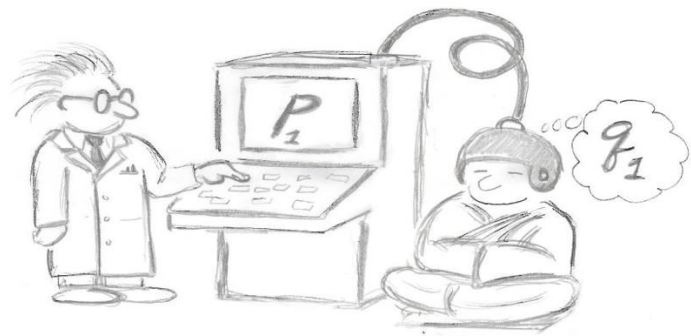

23$^{rd}$ century neuroscientist and thoughtful subject.

**Figure 1: A 23$^{rd}$ century neuroscientist and his thoughtful subject**

In general, we expect that any physical state **p** 'similar enough' to **p₀** will generate the same Yes/No response in the subject. For example, within the theory consisting of semi-classical physics, it seems unlikely that shifting the position of a single oxygen atom 0.1nm in the subjects' brain will produce a different Yes/No answer. Furthermore, any measuring instrument for the states **p** will most likely have some finite error, resulting in a neighborhood of states around **p**. Thus we need to allow for an equivalence class of physical states as well. We will define two states **p** and **p₀** to be in the same *p-equivalence class* **p̃** if they correlate to the same *q-equivalence* class **q̃**.

Thus starting from the initial equivalence class pairing (**p̃₀**, **q̃₀**), the experimenter methodically varies states **p** until the subject first answers 'Yes'; that is, the subject is *certain* to feel to be in a different subjective state. Repeating the same procedure as for the reference pair (**p̃₀**, **q̃₀**) will define a new equivalence pairing (**p̃₁**, **q̃₁**), and so on *ad nauseam* until both the states spaces **P** and **Q** are partitioned into equivalence classes. Assuming the **E₂₃ theory** is successful, the partition will be into disjoint classes, allowing a 1-1 mapping. In this case we might also, for mere convenience, arbitrarily set **C(q̃)=C(p̃)**, where **C(p̃)** is a code chosen by any means deemed convenient to represent the class **p̃**. Note that although a particular theory was harnessed to achieve this coding, the procedure meets the criteria of an operational definition, since it is the subject's judgment that is ultimately used to determine the q-equivalence classes. To recapitulate, the **E₂₃ theory** was used in three ways:

1. As an arbitrary reference system for providing subjective state codes
2. As an external data storage and retrieval system
3. As a theoretical framework for the neuroscientist's sense of understanding.

Note that such an **E₂₃ theory**, in as much as it connects to subjective experiences only through the objectively recordable binary statements Yes/No, can be treated wholly within the realm of Easy Problems. At that point it may be tempting to view such a theory as representing the maximum amount of understanding of subjective experience that is achievable within any scientific program. As noted earlier, this equates to the position that either the Hard Problem has no solution, or subjectivity itself is not a 'real' problem in the first place.

The contrary position is that there exists a corresponding '**H₂₃ theory**' which addresses the Hard Problem. But what could such a theory look like, if by design all reportable information has been exhausted? Note that the above procedure for defining (**p̃**, **q̃**) pairs was carefully constructed to avoid the ambiguity of the subject being *uncertain* as to whether or not a given subjective experience was *exactly equal* to another previous experience. The question was phrased so that the subject gave a positive response only if he is certain that two experiences are different. Here it is important to emphasize the distinction between, on the one hand, a negative finding of not being able to report a difference, and on the other hand, a positive determination of being able to report that there is no difference. While this may seem at first to be a minor technical distinction, in fact it is a crucial one with broad consequences for subjectivity that will now be developed in detail.

## 9. THE NCD PRINCIPLE FOR SUBJECTIVITY

The need for the above distinction lies with another key property I posit for subjectivity: that *one can never determine with certainty that two entire experiences are exactly the same*. We can glibly think of this peculiar fact as the 'No Complete Déjà-vu' principle, or 'NCD' for short. Before going into this particular claim, a sidebar on general claims about subjectivity is in order. Historically within science, appealing to unexamined intuition to establish the truth of a claim has of course been just about the worst way to proceed; the flatness of the earth and the absoluteness of space-time would only be two examples of this. At the same time, any chain of reasoning must have a starting point, and no discussion in any field is possible without appealing to intuition somewhere along the line. But with subjectivity, intuition becomes indispensable in a new way, since its very essence is essentially visceral and private. Thus for most *a priori* claims about the nature of subjectivity, there is *a priori* no objective, formal proof, since we're literally referring to the way we personally experience our own 'raw feelings'. Indeed, the very existence of subjectivity itself as a phenomenon is also only viscerally meaningful to those who have experience themselves.

Nevertheless, some statements about subjectivity can still in some sense be 'experimentally falsified' if their perceived meaning disagrees strongly with those inner visceral senses. Thus the claim by some that subjectivity is not 'real' disagrees strongly enough with many people's direct visceral experience as to effectively refute the claim for them. In the particular case of the NCD claim above; if there are a number of individuals who feel able to tell with *certainty* when two of their experiences in their *entirety* are *exactly* the same, then the NCD principle would stand refuted.

The best to be hoped for to establish such a claim is to give enough illustrative examples that together make a compelling case. Later I will try to bolster and demystify the claim by giving something of an 'explanation' and a broader context for why subjectivity should have this odd property in the first place. For me, the case becomes compelling as soon as I try to imagine claiming any two entire experiences are the same—I find it manifestly impossible. The crucial word is *entire*. While many *aspects* of experience seem to clearly and readily recur (call it 'partial déjà-vu'), such as perceptions of certain sounds, colors, concepts, odors, tastes, fragments of dreams and thoughts, etc., it seems utterly inconceivable to me that I could ever be sure with certainty that *entire experiences* recur. Even if I try to artificially create circumstances which should repeat the same sensations as close as possible – a 'self-experiment' if you will – my uncertainty doesn't diminish. For example, while sitting quietly, look at a some fixed object in a room for a few seconds, then briefly close and then open your eyes, trying to look at the same object in the same way both times. Focus on trying to recreate the same conditions within yourself each time, until you aren't able to report any clear difference between each experience. But at the same time I claim you will notice starkly that you're not able to claim with *certainty* that those experiences *are exactly the same in their entirety.* One can try many more examples, some with physical animation that focuses special attention on the experience: e.g., sitting quietly at a table, clear your mind as much as possible. Then pound your right fist at a fixed angle on the table. Rest, relax, clear your mind for 10 seconds. Repeat and immediately compare your two experiences in their entirety. Again, I claim the same result in this and all cases.

## 10. AN UNCERTAINTY PRINCIPLE FOR SUBJECTIVITY

Once this point has sunk in, the justifiable question is: Why should this be, and what can be so similar about two experiences that you can't report a difference, yet so different that you can't claim they are equal either? Is this just a 'psychological artifact' of our naturally conservative uncertainty in light of the well-known vagueness of memory? But remember there may be little or no uncertainty when comparing individual *aspects* of experience, such as comparing shades of color or sounds, and these must also be recalled from memory. For example, when a particular color is flashed on a screen, then immediately afterwards the same color is flashed again, or when the same musical tone is played twice, there seems to be no such ambiguity. Why can't the same level of certainty be achieved for *entire* experiences?

To use a physical metaphor: ultimately it seems the uncertainty in comparing entire experiences stems from the inability to 'grasp' each entire experience the way we seem to be able to 'grasp' individual aspects 'within them'. For example, I am much better able to 'grasp' a shade of blue than I can 'grasp' the entire experience in which I am, among other things, seeing a shade of blue. Ultimately 'grasping' a particular aspect of an experience means putting it in some sense 'within' another experience that serves to examine that aspect. In the case of aspects, I seem to be able to put a particular aspect of one experience into another experience with complete fidelity. But how do I put an *entire experience* within yet another entire experience for examination? The act of trying to do so already alters that experience to be examined, so that the 'fidelity' of reproducing it is lost, since I can't control how that experience is 'embedded' or in which experience it is 'embedded'. This requires and indeed creates a new experience. Thus, our inability to claim with certainty that any two experiences are equal, far from being some 'psychological artifact', seems rather due to the fact that comparing experiences unavoidably alters them enough to create an irreducible uncertainty. Thus we see the inherent (and frustrating) limitation of any $E_{23}$ **theory** is that, while it incorporates all reportable information of an experience **q**, it is still incapable of adequately addressing the entire subjective experience as an actual experience. Is it principally possible to go beyond this, i.e. to have anything like theory of subjectivity itself?

One theme that runs through the above discussions is that the very act of trying to produce a code for subjective experiences inevitably creates what we can properly think of as a 'measurement error'. Regardless of whether we're satisfied with an $E_{23}$ **theory** in which the subject can only give a Yes/No answer to create equivalence classes, or whether we want to try to code beyond this, in each case the subject needs to reflect on the experience and place it 'within' another, new experience. Therein lays the inescapable rub of any meaningful operational definition of subjective experience: this mental 'measuring' act of introspection itself creates a new and different experience (namely that of reflecting on and coding for the memory of another experience). Thus the very act of coding an experience invariably alters the original state in question in an uncontrollable and unpredictable way. It is clear that this error is irreducible and part of the measurement process itself. In obvious analogy to Heisenberg's principle for quantum mechanics, I'll refer to this situation as an *'uncertainty principle' for subjective states.*

In light of such a strong claim, it should be repeated that just as Heisenberg's principle can't be proven as a theorem within classical physics, but rather introduces fundamentally new physics into the mix, the principles proposed here for subjectivity can't be proven within the context of any purported $E_{23}$ **theory**. As noted earlier, claims about subjectivity rely particularly on this kind of 'non-syllogistic' evidence. Nevertheless, as with Heisenberg's principle, I believe that examinations such as those above make a compelling case for establishing new conceptual tools for subjectivity. In the following we shall see that the analogy with Heisenberg's principle can be pushed quite far, mirroring many of the structural elements of quantum theory.

## 11. MODELING SUBJECTIVE STATES AS INCOMPUTABLE NUMBERS

One approach to modeling subjective states in line with this principle is to treat each state **q** as being inherently unique and irreproducible. This in turn may be readily achieved by modeling any real number **C** representing the state **q** as being *algorithmically incomputable*. Representing subjective state codes $\{C_H\}$ of an $H_{23}$ **theory** as incomputable numbers, as esoteric as it may seem, reflects in a natural and intuitive way some of the properties we have been ascribing to the very peculiar nature of subjectivity, and how subjectivity is related to and differs from *reports* of subjectivity. In particular:

1. <u>No finite algorithm can ever make the determination that two incomputable numbers are exactly equal</u>. From the NCD principle, whereby there is no procedure for establishing the equality of two experiences, this is precisely what we expect of $\{C_H\}$. At the same time, one only needs to examine a finite number of bits to determine when two incomputable numbers are *different* numbers (one need only examine until the first bit where the numbers differ). Again, this is mirrored by the observation that two experiences that differ in at least one aspect (such as seeing a slightly different shade of blue) can be distinguished from another within a finite number of steps.

2. <u>Incomputable numbers are not recursively constructible from any computable set</u>. However, any incomputable number may be approximated to any desired precision with a computable number. This property of $\{C_H\}$ expresses the sense that subjectivity cannot be uniquely 'captured' by reports, however much information is conveyed in them. In a metaphorical sense, subjectivity as it is experienced is 'incompressible'; 'its own shortest description'—often used terminology in describing incomputable numbers.

We may further think of the relationship between an $H_{23}$ **theory**, which takes into account this irreducible uncertainty of its states **q**, and its corresponding $E_{23}$ **theory**, which does not, as analogous to the relationship between a quantum theory and its corresponding classical limit. In particular, any **E₂₃ theory**, whatever other type of incomputability it may or may not contain, is at least free of the above type of incomputability since it is defined in terms of equivalence classes which, by construction, ignore uncertainty in the coding process. Call the codes for these **E₂₃ theory** classes $C_E(\mathbf{q})=C_E(\mathbf{p})$ as defined earlier. We could then imagine writing the codes $C_H(\mathbf{q})$ for an **H₂₃ theory** when $\mathbf{q} \in \mathbf{q}$ as some function $T$ of the corresponding **E₂₃ theory** codes

$$C_H(\mathbf{q}) = T(C_E(\mathbf{q}), \mu),$$

where $\mu$ is an incomputable number and $T$ is a function with the property that $T(C_E(\mathbf{q}), \mu) = C_E(\mathbf{q})$ for the (counterfactual) case $\mu = 0$.

To summarize this heuristic picture then, we may imagine something like this as the relationship between an **E₂₃ theory** (reports of subjectivity) and an **H₂₃ theory** ('subjectivity itself'): Subjective states of **H₂₃ theory** are not constructible from the reports of subjective states of **E₂₃ theory,** but at the same time, **E₂₃ theory** may be said to offer an effective approximation of **H₂₃ theory**.

## 12. A BELL-LIKE INEQUALITY FOR SUBJECTIVE STATES

The analogy with quantum theory can be brought forth at the ontological level as well. Prior to the modern interpretation of quantum mechanics in terms of state spaces/path integrals and probabilities, Heisenberg asked how position and momentum were actually measured as classical variables in experiments; essentially how they are operationally defined. In these thought experiments it became clear that such quantities have an irreducible error that derives from the very act of measurement itself. From the vantage of modern interpretations of quantum mechanics Heisenberg's principle, as originally stated, was developed around a *counterfactual*, namely the convenient fiction that electrons actually have well-defined positions and momenta at all times, and that these follow classical trajectories disturbed by the measurement process. However, the results of EPR-type experiments over the past decades have strongly implied that observables such as electron position and momentum *are merely the outcomes of measurement*, and have no meaningful values until such a measurement takes place. The difference between this ontological view and the competing hidden variables views is generally accepted to be expressed in Bell's inequality, which generally holds for observables **O** having some hidden variable parameterization **O(λ)**. Although this ontological interpretation is still controversial in many circles, the observed violations of Bell's inequality are now usually accepted to imply a violation of local reality.

But be that as it may, we don't have to enter into the long-standing interpretational debate in quantum theory to hypothesize that an analogy with the same sort of ontological framework as might be useful in a theory of subjectivity. As has been shown, the states **q** themselves are not measurable, but rather only their q-equivalence classes **q**. In addition, the incomputability of subjective states codes **C(q)** means *a priori* they can have no 'hidden-variable' parameterization **q(λ)**. Perhaps it is more fruitful to see such states **q**, at least insofar as they are measured to be the q-equivalence classes, to be more the results of our 'mental measurements', i.e. observations, rather than meaningful 'entities' whose ontological status is independent of their 'measurement'.

Indeed it is possible to write an inequality of the form analogous to Bell's theorem as follows. In this presentation we follow a simpler version (Harrison, 1999) more conveniently amenable to the present discussion. Consider a set of objects to be sampled, and three properties **A**, **B** and **C,** that can be measured from each object. Each property can be measured to have the value **0** or **1**. However, for each object only two of the three properties are measured at each sampling. Consider further the functions $F_{AB}$, $F_{BC}$, $F_{AC}$ computed from each such pair-wise measurement as follows:

$$F_{AB} = 1 \quad \text{if } A=1, B=0$$
$$= 0 \quad \text{otherwise}$$

$$F_{BC} = 1 \quad \text{if } B=1, C=0$$
$$= 0 \quad \text{otherwise}$$

$$F_{AC} = 1 \quad \text{if } A=1, C=0$$
$$= 0 \quad \text{otherwise}$$

Let $E(F_{XY})$ be the expectation value $F_{XY}$ over many measurements. If each of the properties **A,B,C** has some 'hidden variable' parameterization **A(λ), B(λ),C(λ)** with some frequency distribution **ρ(λ)**, then it is straightforward to demonstrate the inequality

$$E(F_{AB}) + E(F_{BC}) \geq E(F_{AC}).$$

The 'hidden variable' parameter **λ** specifies that each quantity is 'locally real' in the sense that for a given **λ**, the quantities **A, B** and **C** actually do have well-defined values, whether they are measured or not. If on the other hand, in an EPR-type experimental setup, where **A,B**, and **C** represent, for example, measurement of pairs of entangled spin(1/2) particles along three different axes, respectively, then this inequality is known to be violated, (as is indeed required by quantum mechanics). Hence nature at a microscopic level violates 'local reality' in this sense of hidden variables.

Applying this line of thought to subjectivity, we may imagine our previous 23[rd] century experimental setup with the neuroscientist equipped with an $E_{23}$ **Theory** and generating a specific (**p̄, q̄**) equivalence pair for a subject. Each time the subject is asked to report on two out of three specific aspects **A, B** and **C** from the entire experience. (For example, if the subject reports seeing a certain shade of red, then **A=1,** and **0** if otherwise. When reporting a certain musical tone, then **B=1**, and **0** if otherwise. Similarly, reporting a certain smell gives **C=1,** and **0** if otherwise.) The experimentalist then computes over several repetitions of the equivalence pair (**p̄, q̄**) the expectation values $E(F_{AB})$, $E(F_{BC})$, $E(F_{AC})$ as defined for the functions above. Again, if there exists a parameterization of **A(λ), B(λ),C(λ)**, then the aspects are 'locally real' in the sense of having actual values for each measurement and the expectation values must satisfy the same Bell-type inequality as above. On the other hand, we have seen that the code $C_H(q)$ for subjective state $q$ upon which the aspects ultimately derive from is itself an incomputable number, and hence has no parameterization. Thus it is not *a priori* clear that such a parameterization for **A, B** and **C** must exist in all cases.

Indeed, one can hypothesize that there exist certain neurological test scenarios and some set of aspects **A, B** and **C** such that only two of the three may be reported on per 'mental measuring act'. To take an arbitrary example, it might be that under certain circumstances, the act of focusing and reporting on both a certain color and a certain sound together dominates the mental resources in a manner and to the extent of preventing one from reporting on a certain smell, and so on. If in addition, the reported value of any given aspect varies depending upon which other aspect was also reported on, then the inequality can be readily violated. A concrete realization of this would be the following hypothetical table of measurable outcomes[3]:

If report on **A,B**:  Then either (**A=1, B=1**), or (**A=0, B=0 or 1**)
If report on **B,C**:  Then either (**B=1, C=1**), or (**B=0, C=0 or 1**)
If report on **A,C**:  Then either (**A=1, C=0**), or (**A=0, C=0 or 1**),

which, (as long as **A=1** for at least one **A,C** measurement pair) results in

$$E(F_{AB}) + E(F_{BC}) < E(F_{AC})$$

Such a case violates the condition of 'local reality' in the sense that aspects **A,B,C** cannot be said to have had values until they were 'mentally measured'.

It is worth noting that one can create a plethora of macroscopic test situations that also give the above or similar measurement outcomes, and in this sense violate 'local reality'. For example, a test subject could simply willfully decide to report the above outcomes upon being so queried. Alternatively, a simple logic circuit could readily be built to create the same result. Violations of Bell's inequality in other macroscopic and also in certain cognitive settings have also been discussed previously (Aerts, 2000). However, of particular interest here is whether such an inequality could be violated given repeated measurements with fixed (**p, q**) pairs. Such violations, if observed, would be of singular significance for both the theory of cognitive processing as well as the ontological status of subjectivity and how it relates to physical reality.

---

[3] While the above example is purely hypothetical, a more concrete future experiment might be envisioned in the context of synesthesia, the neurological phenomenon in which stimulation of one sensory or cognitive pathway involuntarily causes the experience of another sensory mode. (e.g., experiencing colors associated with certain numbers and/or sounds.)

## 13. ONTOLOGY OF SUBJECTIVE STATES

It has been argued throughout that the ontological status of subjectivity is problematic. On the one hand from every day speech to psychology to neurobiology it seems quite natural and compelling to speak in an unproblematic way about 'subjective states' **q**. On the other hand, there are profound conceptual reasons to be skeptical of the meaningfulness of this notion. First, we should be suspicious from the fact that, as we have seen, only the q-equivalence classes can be given an operational definition. Even more problematic than operational definitions is the elementary question of just what a subjective *state* – as a representative of the phenomenon of subjectivity – is supposed to mean in the first place. Whatever subjectivity is, it is clearly and profoundly different than literally anything else we're used to talking about. In fact, talking about the ineffable sensing we call subjectivity as if it were an element of a set at all seems at bottom to be a subtle, albeit very hard to avoid, misappropriation of language. Clearly the noun 'subjective state' tries to refer somehow to the first person perspective within us. In saying it that way, we apparently mean it to be something like what we mean by a 'thermodynamic state' or 'neurological state'. But subjectivity's defining essence seems inherently un-capturable with any word-descriptions. In using the noun 'feeling' to try to communicate, we are merely objectifying experience, i.e. creating a noun for it as a false place holder, because that's all we have at our disposal.

The very nature of language as a tool for creating and communicating representations of our subjective experience seems to make it profoundly ill-equipped to deal with the essence of subjectivity itself. Seen in this light, the proposed use of incomputable numbers to represent subjective states might be seen as something of a compromise between the two horns of a profound dilemma: on the one hand, science requires language representations to go about its basic business, yet is called upon to address that which ultimately has no language representation. Incomputable numbers may be seen in this context as sitting in the middle between these two competing requirements, in as much as any representation given is at best a 'place holder' for a number which also has no representation by any finite symbol string.

## 14. CONCLUSIONS AND OUTLOOK

It's clear that $21^{st}$ century science is not even close to a workable $E_{23}$ **theory**, let alone an $H_{23}$ **theory**. Nevertheless, from the present analysis it's possible to at least eliminate some types of theories that are incompatible with any $H_{23}$ **theory**. In particular, it has been argued that a proper operational treatment of subjectivity implies that all subjective states **q** can be represented by incomputable numbers. Thus any deterministic physical system whose states are all necessarily computable (from any computable initial conditions), such as classical physics, is ruled out as subvening on any future $H_{23}$ **theory**. Currently the only widely accepted physical theory that provides a framework for truly nondeterministic processes is, of course, quantum theory. However, this observation is not to imply that quantum theory as currently formulated is a physical foundation of subjectivity. Indeed, incomputable functions are not explicitly contained within the current formalism of quantum theory. Furthermore, it is far from clear what direct, explanatory relevance quantum theory might have. On the one hand, Penrose has argued that the output of at least some mental processes is necessarily nondeterministic (Penrose, 1989). Along these lines it has been hypothesized that such nondeterministic processes, if they exist, might be driven by macroscopically relevant quantum entanglement occurring within the microtubules of neurons within a theoretical framework dubbed Orchestrated Objective Reduction (Hameroff, 1996). On the other hand, Tegmark has argued in detail that the quantum decoherence time scales are orders of magnitude too small for any such processes to be relevant to cognition (Tegmark, 2000). In any event, it's fair to say that there are as yet no clear theoretical or experimental indications of a role for quantum processes in brain functioning. For the moment, it may be best to view the analogy with quantum mechanics as a heuristic one that can provide useful analogies and partial insights into what a future **$H_{23}$ theory** could look like. What the elements and formal structure of such a theory might be, and in particular, whether quantum theory might nevertheless, in principle, be more directly connected conceptually to subjective experience, will be the topic of ongoing investigations.